\documentclass[twocolumn,superscriptaddress,showpacs,floatfix,amsmath,amssymb,aps,pre,longbibliography]{revtex4-1}

\usepackage{color}
\usepackage{hyperref}
\usepackage{graphicx}

\begin{document}
	
\title{Reaction fronts in persistent random walks
with demographic stochasticity}

\author{Davide Vergni}\thanks{Corresponding author}\email{davide.vergni@cnr.it}
\affiliation{Institute for Applied Mathematics ``Mauro Picone'', National Research Council of Italy, 00185 Rome, Italy}

\author{Stefano Berti}
\affiliation{Universit\'e de Lille, Unit\'e de M\'ecanique de Lille, UML EA 7512, F-59000 Lille, France}

\author{Angelo Vulpiani}\affiliation{Dipartimento di Fisica, ``Sapienza'' Universit\`a di Roma, p.le A. Moro
2, 00185 Roma, Italy}

\author{Massimo Cencini}\thanks{Corresponding author}\email{massimo.cencini@cnr.it}
\affiliation{Istituto dei Sistemi
	Complessi, CNR, via dei Taurini 19, 00185 Rome, Italy}


\date{\today}

\begin{abstract}
Standard Reaction-Diffusion (RD) systems are characterized by infinite
velocities and no persistence in the movement of individuals, two
conditions that are violated when considering living organisms.  Here
we consider a discrete particle model in which individuals move
following a persistent random walk with finite speed and grow with
logistic dynamics.  We show that when the number of individuals is
very large, the individual-based model is well described by the
continuous Reactive Cattaneo Equation (RCE), but for smaller values of
the carrying capacity important finite-population effects arise.  The
effects of fluctuations on the propagation speed are investigated both
considering the RCE with a cutoff in the reaction term and by means of
numerical simulations of the individual-based model. Finally, a more
general L\'evy walk process for the transport of individuals is
examined and an expression for the front speed of the resulting
traveling wave is proposed.

\end{abstract}
\maketitle

\section{Introduction}
The spreading of reactive quantities, such as, e.g., biological
populations or chemicals, is often conveniently modeled by means of
reaction-diffusion (RD) equations. This approach finds application in
fields as diverse as combustion~\cite{clavin1994premixed},
genetics~\cite{fisher1937wave,aronson1975nonlinear,korolev2010genetic}
epidemics' spreading~\cite{murray1989} and ecology~\cite{okubo}.  By
representing transport through standard diffusion, RD descriptions
allow for the instantaneous spreading of the transported species over
arbitrarily large distances from their original location (albeit with
a very small probability).  From the point of view of the individual
reactive entities these features translate into motions with infinite
velocity and no inertia.
These assumptions are not realistic and seem particularly problematic
in biology
\cite{skellam1951random,stinner1983dispersal,turchin1989beyond,holmes1993diffusion}. In
fact, all organisms displace themselves at a finite velocity, with
persistent movements (i.e. with some inertia to change velocity), at
least over short time intervals
\cite{skellam1951random,okubo,codling2008random,garcia2011random,mendez2014random}.

Using a continuous field description, suitable generalizations of RD
models have been proposed to remedy the above mentioned unphysical
features in different contexts (see
\cite{holmes1993diffusion,hadeler1994travelling,lemarchand1998perturbation,horsthemke1999fisher,
  galenko2000hyperbolic}, and \cite{fortmendez2002} for a review). In
the framework of population dynamics such theoretical approaches have
proven useful to interpret previously controversial data about the
spread of virus infections \cite{fort2002time} and human population
invasions \cite{fort2003population}.

Here, we consider a system of individuals that move in a correlated
way with a finite speed, and that reproduce (or die) with prescribed
reaction kinetics.  Our main goal is to gain insights into the way the
population spreads in space under the combined action of the
generalized diffusive process and reaction, as well as to assess the
role of demographic stochasticity, namely the fluctuations in the
number of individuals associated with the discrete and stochastic
nature of the population, whose importance is well known
\cite{durrett1994importance}. We will particularly focus on the speed
of invasion into an unoccupied environment, starting from a localized
source, in the different dynamical regimes of the system.

As for the generalized diffusive dynamics we consider a simple model
in which the particles travel for a certain time maintaining their
(finite) velocity and then change it randomly. This kind of model is
rather flexible as, properly choosing the distribution of travel
durations, it can reproduce several transport
processes, including L\'evy walks \cite{zaburdaev2015levy}.  For the
reaction, we consider a logistic growth model, which is the simplest
possible mathematical description accounting for reproduction and
death due to competition for resources, and it also applies to simple
autocatalytic chemical reactions \cite{murray1989}. This choice is
further motivated by the fact that in the context of RD processes of
the pulled kind
\cite{vansaarlos}, this corresponds to the prototypical
Fisher-Kolmogorov-Petrovskii-Piskunov (FKPP) model
\cite{fisher1937wave,kolmogorov1937}, for which the effects of
discreteness on propagating fronts (i.e. traveling wave solutions)
have been shown to be well captured by the introduction of a
small-density cutoff in the reaction term \cite{brunet1997}.

The article is organized as follows.  In Sec. \ref{sec:model} we
introduce the stochastic model for the transport and reaction dynamics
of particles. In Sec. \ref{sec:cattaneo} we investigate the continuous
limit of the particle model and show that it corresponds to the
reaction Cattaneo equation (RCE)
\cite{patlak1953random,codling2008random}. We first discuss front
propagation in the RCE for both small and large reaction rates
corresponding to a RD-like and to a ballistic regime of propagation,
and then examine the effect of truncating the reaction term at small
densities in both regimes so as to mimic, within the continuum
framework, the effect of demographic stochasticity \cite{brunet1997}.
In Sec. \ref{sec:results} we numerically study the stochastic particle
model introduced in Sec. \ref{sec:model} to quantify the demographic
stochasticity effects and compare it with the continuum
description. We will see that the phenomenology of the
individual-based system in the ballistic regime is richer than in the
continuous description. Finally, in Sec. \ref{sec:extension} we
present a preliminary study of the particle model in which the
transport process is generalized to a L\'evy walk, with particles'
velocities persisting for random durations distributed 
according to a fat-tailed probability density function. 
Discussions and conclusions are presented in
Sec. \ref{sec:concl}. 
In Appendix~\ref{app:derrida} we generalize the derivation 
of Ref.~\cite{brunet1997} to the case of the RCE 
with a cutoff. In Appendix~\ref{app:logistic} we 
present an exact solution of
the stochastic logistic dynamics in the absence of transport processes.

\section{Model\label{sec:model}}
We consider a stochastic model of a population of individuals that
perform a persistent random walk and that reproduce/die with density
dependent rates.  For simplicity, we consider a one-dimensional
system. In the following we separately describe how individuals move
in space and their reaction dynamics.

\paragraph*{Particle transport} Each individual moves independently from the
others by maintaining its velocity $v$, extracted with probability
$p(v)dv$, for a walk lasting a time $T$, which can also be a random
variable, independent of $v$, chosen with probability
$P(T)dT$. Assuming that $\langle v^2 \rangle$ and $\langle T^2
\rangle$ are finite and that $\langle v\rangle=0$, one has that at
short times the motion is ballistic while asymptotically it becomes
diffusive. The diffusion coefficient may be obtained with a simple
argument as follows \cite{vulpio}. Let us denote with $t_i=
\sum_{k=1}^{i} T_k$ the sequence of times at which a new velocity,
$v_i$, is chosen and let $w_t$ be the number of walks up to time $t$.
Then the position, $x(t)$, of the particle at time $t$ can be written
as $x(t)=\sum_{i=1}^{w_t}v_iT_i$, where $x(0)=0$ without loss of
generality.  Since the random variables are all independent, for the
dispersion of the position we can write
\begin{equation}
  \langle  x(t)^2\rangle=
  \left \langle  \sum_{i=1}^{w_t}v^2_iT^2_i \right \rangle =
  \langle v^2\rangle\langle T^2\rangle w_t = 
  \langle v^2\rangle \frac{\langle T^2\rangle}{\langle T \rangle} t\,, 
  \label{eq:x2}
\end{equation}
where we used that $\sum_{i=1}^{w_t} T_i = t = w_t \langle T\rangle$,
which holds for $w_t\gg 1$.  The above equation displays a diffusive
behavior $\langle x^2(t)\rangle=2Dt$ with diffusion coefficient
\begin{equation}
  D=\frac{\langle v^2\rangle}{2}\frac{\langle T^2\rangle}{\langle T \rangle}\,.
  \label{eq:D}
\end{equation}

In the present model the velocity distribution is assumed to be $p(v)=
\frac12 \delta(v+u)+\frac12 \delta(u-v)$, while, for the walk duration
we take $P(T)=\delta(T-1)$, i.e.  the walk time is fixed to
$T=1$. With these choices Eq.~(\ref{eq:D}) implies that the diffusion
coefficient is equal to $D=u^2/2$.  We stress that the results we are
going to present are robust and independent of the specific choices of
$P(T)$ and $p(v)$ (as confirmed by tests done with exponentially
distributed times and Gaussian distributed velocities, not shown)
provided the motion is asymptotically diffusive, i.e. when $T$ and $v$
have finite variance and there is no correlation between them. In
Sec.~\ref{sec:extension} we will consider a more general distribution
for the time duration to account for the possibility of L\'evy walks.

\paragraph*{Reaction dynamics}
When dealing with a particle description, in principle, one has to
consider the reaction among particles which are within a certain
interaction distance, $R$. This kind of approach requires to follow
the particles and, at each time step, to perform the reaction for all
particles falling inside the interaction distance. This is quite
expensive from a computational point of view. To ease the computation
we used a modification of the approach proposed in
\cite{pigolotti2,pigolotti1}.  The domain of size $L$ is divided in
$L/R$ bins of size $R$. The number of particles $n_i(t)$, 
whose positions at time $t$ fall in the $i$-th bin ($i= 1,\ldots,L/R$)
is evolved according to the rate equations:
\begin{eqnarray}
  n_i(t+dt) &\to& n_i(t)+1 \quad w.p. \quad rn_i(t)\,dt \label{eq:birth}\\
  n_i(t+dt) &\to& n_i(t)-1 \quad w.p. \quad r{n_i(t)^2}/{(NR)}\, dt  \label{eq:death}\\
  n_i(t+dt) &\to& n_i(t)\phantom{+1\;}   \quad  \mathrm{otherwise}
\label{eq:rate3}
\end{eqnarray}
where $N$ is the density of carrying capacity, i.e. in each bin the
expected number of individuals is $N_p=NR$.  Neglecting particle
migration in and out of the bin, the above rates ensure that
$dn_i/dt=rn_i(1-n_i/N_p)$ plus a zero average stochastic term,
i.e. they reproduce the logistic growth dynamics. From an algorithmic
point of view, birth (\ref{eq:birth}) and death (\ref{eq:death}) events
are implemented by choosing a random individual among the $n_i$
present in the $i$-th bin and cloning or removing it, respectively. In
the case of birth, the cloned individual is initialized at the same
position of the parent with velocity $v$ and walk time $T$ randomly
extracted according to the chosen probability distributions.

In our simulations we initialize the population by seeding ten bins
around the center of the domain ($L/2$) with $N_p/2$ particles
uniformly distributed within each bin.  The numerical integration is
carried on until one particle reaches a boundary (at $x=0$ or $x=L$)
so as to avoid boundary effects. The time step $dt$
has to be chosen in such a way that the probabilities on the r.h.s. of
(\ref{eq:birth}-\ref{eq:rate3}) are much smaller than $1$.  As for the
system size, we used $L=(10^5-10^6) \,R$ in order to ensure reliable
estimates of the propagation speed. Finally, we fixed $R=0.1$ and
checked that all the results are not influenced by this choice.

\section{Continuum limit: the Reactive Cattaneo Equation
(RCE)\label{sec:cattaneo}}
When the population is very large, i.e. in the limit of large carrying
capacity $N\to \infty$, the stochastic model presented in the previous
section is expected to follow the reactive Cattaneo equation (RCE)
\cite{holmes1993diffusion,fortmendez2002}.  The RCE can be obtained
starting from different microscopic models as reviewed in
\cite{fortmendez2002}, and following this paper we specialize the
derivation to our model. Denoting with $n(x,t)$ the density of
particles~\footnote{If $x$ corresponds to the $i$th bin, i.e. $x=Ri$,
  $n(x,t)$ is defined by $n(x=iR,t)=n_i(t)/(NR)$ in the limit $N\to
  \infty$.} at time $t$ and at position $x$, we can write
\begin{eqnarray}
n(x,t+T)&=&\frac{1}{2}n(x-uT,t)+\frac{1}{2}n(x+uT,t) +\nonumber \\ && [n(x,t+T)-n(x,t)]_{r}\,, \label{eq:density1}
\end{eqnarray}
where the first two terms account for the transport process and the
last term stands for the variation of the number of particles due to
the reaction. At long times $t\gg T$ and large distances $x\gg uT$,
upon expanding (\ref{eq:density1}) up to second order one obtains the
following equation
\begin{equation}
\frac{T}{2} \partial^{2}_{t} n + \partial_t n = u^2\frac{T}{2} \partial_{x}^{2} n + \frac{T}{2}\partial_t F(n) +F(n)\,, 
\label{eq:cattaneo1}
\end{equation}
where $F(n)$ stands for $[\partial_t n]_{r}$.
Equation~(\ref{eq:cattaneo1}) can then be rewritten in  
the standard form of the RCE \cite{fortmendez2002}:
\begin{equation}
\tau \partial^{2}_{t} n + (1-\tau F^{\prime}(n))\partial_t n = D \partial_{x}^{2} n + F(n)\,, 
\label{eq:cattaneo}
\end{equation}
where $\tau=T/2$ and $D=\tau u^2$; $F^{\prime}$ denotes the first
derivative with respect to the argument.  Fixing $T=1$ as in our model
$\tau=1/2$ and $D=u^2/2$, consistently with (\ref{eq:D}), and given
the reaction kinetics (\ref{eq:birth}-\ref{eq:rate3}), the reaction
term $F(n)$ has the usual logistic form $F(n)=rn(1-n)$.

The RCE has been considered in several previous studies (see, e.g.,
\cite{holmes1993diffusion,horsthemke1999fisher,fortmendez2002}).  It
is not difficult to derive the expression for the asymptotic front
speed (see for example \cite{fortmendez2002}.  Using arguments similar
to those of Brunet and Derrida \cite{brunet1997} it is also possible
(as shown in Appendix~\ref{app:derrida})
to analytically investigate how the front speed changes in the
presence of a reaction cutoff mimicking the effect of population
discreteness.  Both these aspects will be considered in the following
subsections, in particular, the latter will be the guideline for
interpreting the results of simulations of the stochastic model
introduced in Sec.~\ref{sec:model}.

To ease the forthcoming analysis it is useful to rewrite
(\ref{eq:cattaneo}) in a non-dimensional form by introducing
$\tilde{x}=x \sqrt{r/D}$ and $\tilde{t}=rt$ where $r\equiv
F^{\prime}(0)$.  In these variables (\ref{eq:cattaneo}) reads (tildes
suppressed):
\begin{equation}
a \partial^{2}_{t} n
+ (1-a f^{\prime}(n))\partial_t n = \partial^2_{x} n + f(n) 
\label{eq:cattaneo2}
\end{equation}
where $f(n)=F(n)/r=n(1-n)$ and $a=r\tau$. Notice that for $a=0$ the
above equation recovers the standard FKPP model $\partial_t n=
\partial^2_{x} n +f(n)$ \cite{kolmogorov1937}.
\subsection{Front speed from linear analysis}
The basic phenomenology of Eq.~(\ref{eq:cattaneo2}) can be understood
assuming a traveling wave solution $n(x,t)=h(z)$, with $z=x-v_f t$,
and linearizing around $h\approx 0$ (see also \cite{fortmendez2002}),
which is the standard procedure to investigate pulled fronts
\cite{vansaarlos}.  The linearization of Eq.~(\ref{eq:cattaneo2})
yields
\begin{equation}
(1-av_f^2) h^{''}+v_f(1-a)h^{'}+h=0\,.
\label{eq:linear}
\end{equation}
Assuming an exponential leading edge $h(z)\sim \exp(-\lambda z)$ 
the characteristic equation is obtained and its solution
provides the dispersion relation
\begin{equation}
v_f(\lambda)
= \frac{-(1-a)+\sqrt{(1+a)^2+4a\lambda^2}}{2a\lambda}\,.
\label{eq:disprel}
\end{equation}
The plus sign in front of the square root is due to our choice
$z=x-v_f t$, with $v_f>0,$ corresponding to left-to-right propagation.
Notice that Eq.~(\ref{eq:disprel}) has an asymptote
$v_f(\lambda)=1/\sqrt{a}$ for $\lambda\to \infty$ corresponding to the
ballistic speed $v_f=u$ in physical units. This is physically sound,
as the front speed cannot exceed the particle's velocity $u$.  For
$a<1$, $v_f(\lambda)$ has a minimum
\begin{equation}
  v_0=v(\lambda_0)= \frac{2}{1+a}\,\,\,\,\mbox{for}\,\,\,\,
  \lambda_0=\frac{1+a}{1-a}
  \label{eq:v0}
\end{equation}
that, for sufficiently localized initial conditions (i.e. decaying
faster than exponentially, as usual in the FKPP problem
\cite{vansaarlos}), is the selected speed of the traveling front.  The
minimum disappears in favor of the asymptote $1/\sqrt{a}$ when $a\geq
1$.

Summarizing, in physical units the front speed is given by
\begin{equation}
  v_0=\left\{
  \begin{array}{ll}
2u \frac{\sqrt{r\tau}}{1+r\tau} & \mathrm{if} \quad r\tau<1\\
u & \mathrm{if} \quad r\tau\geq 1
  \end{array}
  \right.
  \label{eq:vfront}
\end{equation}
Notice that for $a=r\tau \to 0$ one gets back the FKPP result
$v_f${\tiny $^{\mathrm{FKPP}}$}
$=2\sqrt{Dr}=2u\sqrt{r\tau}$, while for $r\tau>1$
$v_0<v_f${\tiny $^{\mathrm{FKPP}}$} always.  
When $r\tau\geq 1$ the minimal speed from
the dispersion relation is always realized at $v_f=u$ with $\lambda
\to
\infty$ so that the front is expected to evolve ballistically with the
intrinsic speed of the particles and with a very steep (more than
exponential) front.  To the best of our knowledge
Eq.~(\ref{eq:vfront}) was first derived in \cite{holmes1993diffusion}
using a different method, the procedure here followed is rather
standard for the FKPP equation~\cite{vansaarlos,Lopez} and has been
already used for the RCE~\cite{fortmendez2002}.

\subsection{Effects of a cutoff on the front speed\label{sec:cutoff}}
Following the approach of Brunet and Derrida \cite{brunet1997}, let us
now modify (\ref{eq:cattaneo2}) by assuming that the reaction takes
place only if $n>\epsilon$, with $\epsilon$ a given threshold
mimicking the effect of discreteness of the population. This amounts
to replacing the reaction term with $f_\epsilon(n)=f(n)c_\epsilon(n)$
where $c_\epsilon(n) \to 0$ when $n\leq
\epsilon$. Following \cite{brunet1997} we take $c_\epsilon(n)=
\Theta(n-\epsilon)$, where $\Theta$ is the Heaviside step function.  
We must then distinguish two cases depending on whether $a=r\tau$
is smaller or larger than $1$.

When $a=r\tau<1$, the RCE recovers the basic 
phenomenology of the FKPP dynamics and generalizing
the derivation of Ref.~\cite{brunet1997} (see Appendix~\ref{app:derrida}), 
one finds that the front speed, $v_f$, is given by
\begin{equation}
v_f= v_0 - \frac{1}{2} v^{\prime\prime}(\lambda_0)\frac{\pi^2 \lambda_0^2}{(\log \epsilon)^2}
\label{eq:prediction}
\end{equation}
where $v^{\prime\prime}$ denotes the second derivative of the dispersion
relation (\ref{eq:disprel}), $\log$ is the natural logarithm and 
$\lambda_0$ and $v_0$  are given in
Eq.~(\ref{eq:v0}).

\begin{figure}[htb!]
\centering
\includegraphics[width=0.95\columnwidth]{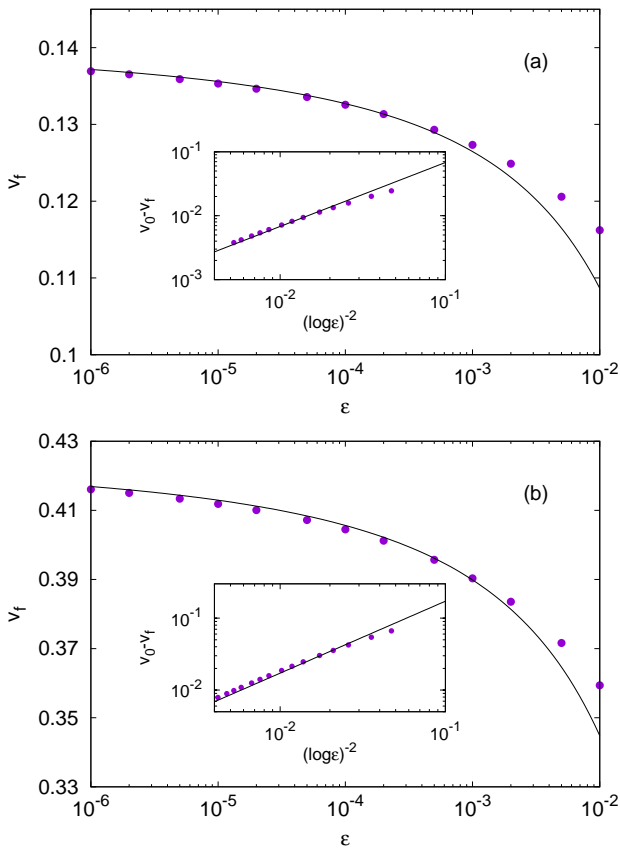}
\vspace{-.4truecm}
\caption{(Color online) Front speed $v_f$ vs the cutoff value $\epsilon$ 
for the RCE with $\tau=1/2$, $u=1$ and (a) $r=0.01$ and (b) $r=0.1$.
The solid black curves show the theoretical prediction
(\ref{eq:prediction}).  In particular, the constant
$A=v^{\prime\prime}(\lambda_0)(\pi^2 \lambda_0^2/2)$ in front of
$(\log\epsilon)^{-2}$ takes the values $A=0.68$ and $1.72$ for
$r=0.01$ and $r=0.1$, respectively.  The insets show the same data
plotting $v_0-v_f$ against $(\log\epsilon)^{-2}$ to highlight the
logarithmic correction, the solid lines are again the
theoretical predictions.
\label{fig:cattaneoFKPP}}
\end{figure}

The validity of (\ref{eq:prediction}) is confirmed in
Fig.~\ref{fig:cattaneoFKPP} where we show the results from numerical
simulations of (\ref{eq:cattaneo}) for $a=r\tau=0.005$ and $0.05$ as a
function of the cutoff $\epsilon$.

\begin{figure}[t!]
\centering
\includegraphics[width=\columnwidth]{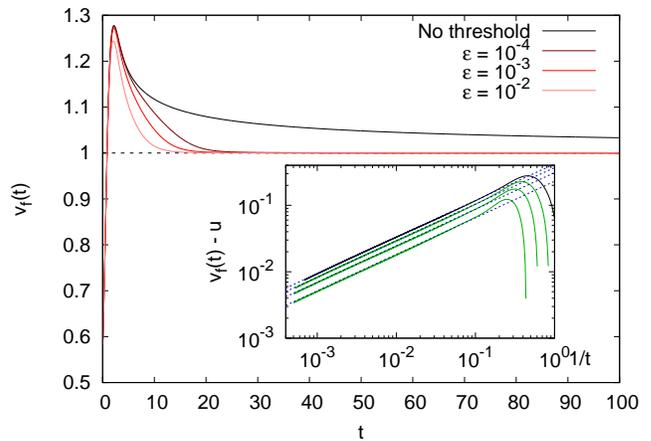}
\vspace{-.4truecm}
\caption{(Color online) Instantaneous front speed $v_f(t)$, as defined 
  in main text, vs time for the RCE with $\tau=u=1$, $r=2$ and four
  different values of the cutoff.  The dashed horizontal line stands
  for the ballistic speed $u=1$. The inset shows $v_f(t)-u$ vs $1/t$,
  to illustrate the trend toward the ballistic speed in the absence
  of cutoff (from top to bottom $a=r\tau=2.00,1.75,1.50,1.25$).  The
  top black curve corresponds to that of the main panel. The dashed
  lines display the power law behavior $t^{-0.55}$. Deviations from
  the behavior $t^{-0.55}$ are observed for $a=r\tau\sim 1$ (not
  shown).  \label{fig:cattaneoBallistic}}
\end{figure}

Conversely, when $a=r\tau>1$, the approach of
Ref.~\cite{brunet1997} cannot be used as the linearized treatment
becomes meaningless.  However, heuristically we can expect that since
in this limit the front evolves ballistically the velocity becomes
independent of the cutoff and equal to the maximal allowed velocity
$v_f=u$. Figure~\ref{fig:cattaneoBallistic} displays the numerically
observed behavior of the front speed as a function of time for the RCE
with different values of the cutoff $\epsilon$, when $a=r\tau=2$.
As one can see, at long times the front speed approaches the
asymptotic speed $u$ independently of the value of the cutoff.  It is
worth noting that the asymptotic speed is approached from greater
values.  Moreover, while with the cutoff the limiting value $v_f=u$ is
reached rather quickly, in its absence the convergence is rather slow.
Indeed, as shown in the inset of Fig.~\ref{fig:cattaneoBallistic}, we
found approximately $v_f(t) -u \sim t^{-0.55}$ for $a=r\tau$
sufficiently large.  This behavior is quite different from
what happens in the FKPP case, in
which one has, at leading order, 
$v_f(t)-u \sim -t^{-1}$
\cite{vansaarlos,berestycki2018new}. At present we have no explanation
for the value $0.55$ of the exponent. This may deserve further
investigation but it goes beyond the scope of the present work.

We conclude this section providing some details on the numerical
integration of the RCE. In order to have a stable and robust numerical
scheme we transformed the original RCE in a system of two first order
partial differential equations whose dynamics follow the
characteristic functions of the linear wave equation associated to the
RCE~\cite{aregba2016time}. Then we used a Roe's first-order upwind
scheme \cite{roe1987upwind} for the numerical integration of the PDE
system. As for the initial condition, we have chosen it to be
localized around the center of the
system. We measured the instantaneous front speed
as $v_f(t)= \frac{1}{2} \partial_t
\int_0^L n(x,t){\mathrm d}x$, which provides an estimate of the bulk
reaction speed~\cite{constantin2000bulk}. The factor
$1/2$ is due to the front propagating in both directions.  The
limiting speed is obtained by extrapolating the behavior of $v_f(t)$
for long times. The simulation stops whenever $n(0,t)$ or $n(L,t)$ is
different from zero to avoid boundary effects.

\section{Effects of demographic stochasticity\label{sec:results}}
In this section we consider the stochastic individual-based model
introduced in Section~\ref{sec:model} in order to study how changes in
the carrying capacity, and thus the fluctuations of the number of
individuals, influence the front speed, having as a guiding line the
results obtained in the continuum limit (Sec. \ref{sec:cattaneo}).

Before starting with the analysis, a comment about the definition of
the front speed in the discrete case is in order.  A first and
natural definition can be given in terms of the growth rate of the
total number of particles in the systems $N_T(t)$ that, within our
model, corresponds to the sum of the number of particles in all the
bins in which the domain is discretized, i.e. $N_T(t)=\sum_{i=1}^{L/R}
n_i(t)$.  By analogy with the definition of the front speed given at
the end of the previous section in the case of a continuous
system~\cite{constantin2000bulk}, we can define the instantaneous
front speed as
\begin{equation}
v^b_f(t) = \frac{1}{2N} \frac{dN_T(t)}{dt}\,,
\label{eq:stdvf}
\end{equation}
which expresses the velocity as a bulk property.
The word bulk refers to the fact that due to the space
average we capture only the large scale properties referring to the
whole particle system.
The factor $1/2$ accounts
again for the fact that propagation occurs in both directions and we
recall that $N=N_p/R$ is the density of carrying capacity, where $R$
is the bin size and $N_p$ is the carrying capacity in a bin.

However, it is also possible to define the front speed in terms of the
positions of the extremal particles. Denoting with $x_{m/M}(t)$ the
position of the left/rightmost particle, we can define the extremal
velocity as
\begin{equation}
v_f^{e}(t) = \frac{x_{M}(t)-x_{m}(t)}{2t}\,.
\label{eq:mmvf}
\end{equation}
This definition does not probe a bulk property of the traveling front
but only concerns the behavior of its edges.  In both cases, the
asymptotic (long time) front speed, which is the quantity we are
interested in, can be obtained extrapolating the constant behavior in
the limit of long times, i.e. $v_f=\lim_{t\to \infty}
v_f(t)$. Numerically this is done by means of a linear fit of the long
time behavior of $N_T(t)/(2N)$ and $(x_{M}(t)-x_{m}(t))/2$,
respectively.  As we will see the two definitions may not always lead
to the same asymptotic front speed. The above result is at odds with the
continuous case, where the bulk and extremal speeds always
coincide. In that case the former is defined as at the end of the
previous section, while the latter can be defined by introducing a
threshold value on the particle density.

\begin{figure}[t!]
\centerline{\includegraphics[width=0.95\columnwidth]{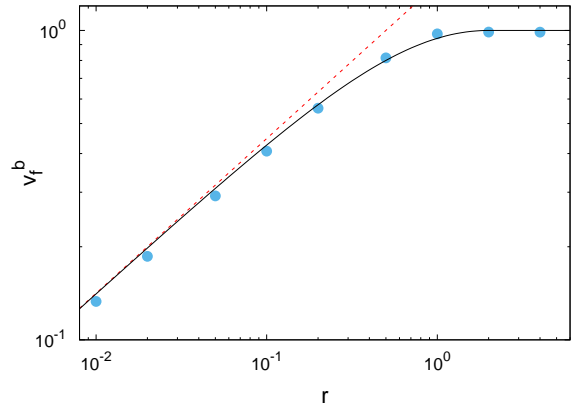}}
\vspace{-.4truecm}
\caption{(Color online) Bulk front speed $v^b_f$ vs $r$ for the 
stochastic particle model with
 $N_p=100$ and $u=1$, compared with Eq.~(\ref{eq:vfront}) (solid line) and 
the FKPP front speed $v_f^{^\mathrm{FKPP}}\!\!=2 \sqrt{rD}$
 (dashed line).}
\label{fig:velvsr}
\end{figure}
Let us now discuss the main numerical results. First of all, we
measured the asymptotic front speed upon fixing the carrying capacity
and varying the reaction rate $r$, to test whether the continuum-limit
prediction (\ref{eq:vfront}) catches the behavior of the
individual-based model. In Figure \ref{fig:velvsr} we show the bulk
front speed, $v^b_f$, obtained using the definition (\ref{eq:stdvf})
(indistinguishable results are obtained using Eq.~(\ref{eq:mmvf})).
As one can see, Eq.~(\ref{eq:vfront}) well captures the behavior of
$v^b_f$, confirming that the RCE indeed provides the continuum limit
of the system under consideration.  The front speed of the FKPP model
also appears to be a good approximation for $r\tau\ll 1$ (see the
dashed line in the plot).  However, small deviations (here hidden by
the scale of the graph) are present. These are due to the fluctuations
of the number of individuals that are unavoidable in the discrete
case.  In the following we study in detail how such fluctuations
affect the front speed. Knowing from the study of the RCE with a
cutoff that the two regimes $r\tau<1$ and $r\tau>1$ are different we
will discuss them separately.

\subsection{Low reaction rates\label{sec:lowrates}}

\begin{figure}[t!]
\centering
\includegraphics[width=0.95\columnwidth]{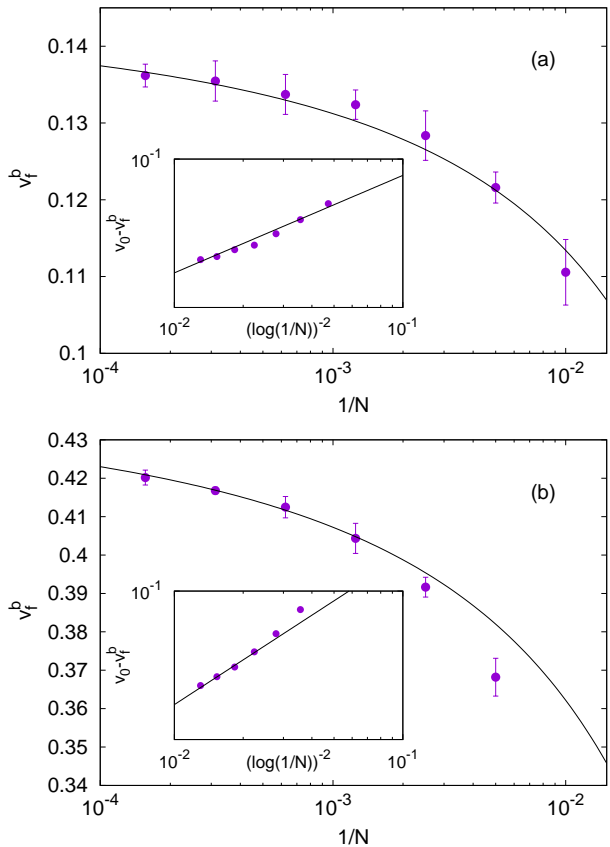}
\vspace{-.4truecm}
\protect
\caption{(Color online) Front speed $v^b_f$ vs $1/N = R/N_p$ as obtained 
  from simulations of the stochastic model with $u=1$ and (a) $r=0.01$
  and (b) $r=0.1$.  The solid curve is obtained by fitting the
  expression $v^b_f=v_0-A (\log(1/N))^{-2}$ where we fixed $A$
  according to Eq.~(\ref{eq:prediction}) and fitted $v_0$; the latter
  resulted to be $4\%$ higher than the continuum-limit value
  (\ref{eq:vfront}).  The insets show the same data plotting
  $v_0-v^b_f$ against $(\log(1/N))^{-2}$ to highlight the logarithmic
  correction. We show the average over $10$ simulations with
  different realizations of the noise, and error bars represent
  the maximal deviation from the mean.}
\label{fig:stochr0.01}
\end{figure}

For low reaction rates, $r\tau<1$, as discussed in
Sec.~\ref{sec:cattaneo}, the RCE behaves essentially as a standard RD
system and the effect of a cutoff, $\epsilon$, on the reaction is well
described by the results of Brunet and Derrida~\cite{brunet1997} (see
Fig.~\ref{fig:cattaneoFKPP}), originally derived for FKPP-like
dynamics. Hence, we should expect that changing the carrying capacity
in the stochastic model should have an effect similar to that of
varying the cutoff in the RCE and, thus, that the front speed should
behave according to the prediction (\ref{eq:prediction}) with
$\epsilon \sim 1/N = R/N_p$. This is confirmed in Fig.~\ref{fig:stochr0.01} 
that shows the bulk front speed, $v^b_f$, as
a function of $1/N$ for the same reaction rates as those chosen for
the RCE (Fig.~\ref{fig:cattaneoFKPP}). The prediction
(\ref{eq:prediction}) is quantitatively well verified but for a small
difference in the value of the $N\to \infty$ velocity, indeed the
fitted value of the velocity differs from the theoretical value
(\ref{eq:vfront}) by $4\%$.
Equivalent results are obtained using the extremal velocity 
(\ref{eq:mmvf}), at least for large $N$.

\subsection{High reaction rates\label{sec:highrates}}
\begin{figure}[h!]
\centering
  \includegraphics[width=0.95\columnwidth]{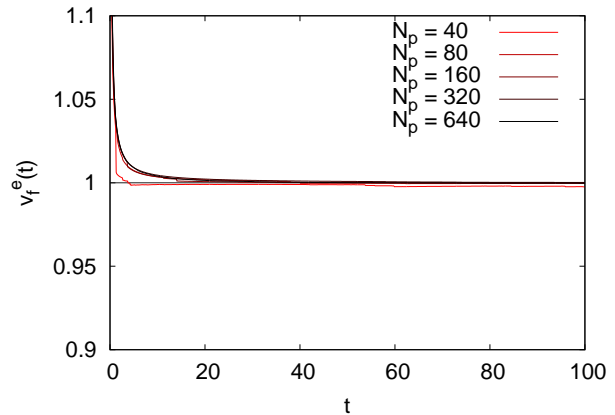}
\vspace{-.4truecm}
\caption{(Color online) Instantaneous front speed $v_f^{e}(t)$ 
  as a function of time
  computed via Eq.~(\ref{eq:mmvf}) for different values of $N_p$ for
  $r=2$ and $u=1$, as obtained by simulation of the stochastic
  individual-based model. Each curve corresponds to an average over
  $10$ realizations of the process.}
\label{fig:vfmaxmin}
\end{figure}

We now turn to the case $r\tau>1$.  In this regime, for the continuous
model, the front speed is unaffected by a cutoff in the reaction term 
(Fig.~\ref{fig:cattaneoBallistic}).  For the
individual-based model, instead, simulations show that the effect of
fluctuations on the front speed depends on the definition adopted for
$v_f$. The bulk speed (\ref{eq:stdvf}) displays a dependence on the
carrying capacity $N_p$, while the speed based on the evolution of the
front edges (\ref{eq:mmvf}) is consistent with the results of the
continuum limit.  The behavior of the latter is shown in
Fig.~\ref{fig:vfmaxmin}, where the time evolution of $v^e_f(t)$ is
shown.  The qualitative features are indeed akin to those of
Fig.~\ref{fig:cattaneoBallistic}: the rightmost and leftmost front
edges asymptotically move ballistically into the unoccupied regions
ahead of them.

\begin{figure}[t!]
\centering
  \includegraphics[width=0.95\columnwidth]{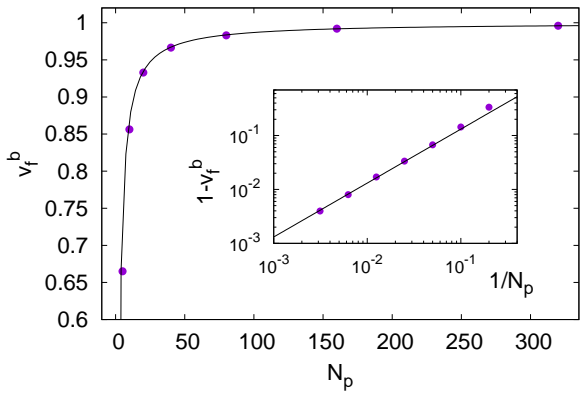}
\vspace{-.4truecm}
\caption{(Color online) Bulk front speed $v^b_f$ vs $N_p$ for $r=4$ 
and $u=1$ obtained by simulations of the particle 
model (symbols).  The solid curve shows the functional form
(\ref{eq:vergnifit}) with $C=1.31$ as obtained by a best
fit procedure.  The agreement between (\ref{eq:vergnifit}) and the
data is very good, as also shown in the inset, where $u-v^b_f$ is
plotted against $1/N_p$ for both the numerical data (symbol) and the
behavior (\ref{eq:vergnifit}) (solid line).  \label{fig:soprasoglia}}
\end{figure}

Conversely, as shown in Fig.~\ref{fig:soprasoglia}, the bulk front
speed, $v_f^b$, obtained as the long time limit behavior of
(\ref{eq:stdvf}) displays a non-trivial dependence of the front speed
on the carrying capacity $N_p$.  In particular, we found 
\begin{equation}
v^b_f = u\left(1-\frac{C}{N_p}\right)
\label{eq:vergnifit}
\end{equation}
to hold, with a high degree of accuracy for large $N_p$.

Clearly we cannot use the continuum theory to explain such a behavior,
and the possible explanation must rely on the particle nature of the
system, in particular on the stochastic nature of the reaction term
that could impact the effective value of the carrying capacity in the
bulk. Indeed, at long times the total number of particles
is expected to evolve as $N_T(t)= 2 \langle n(N_p) \rangle v^{e}_f
t/R$. In other words, at long times $N_T$ will be simply the number of
invaded bins $2v^{e}_f t/R$ (we thus used the definition based on the
extremal bins, neglecting the fact that they may have not 
reached the maximal capacity yet, which is a good approximation at long times)
times the average number of individuals in each bin $\langle n(N_p)
\rangle$. Now, using that $v^{e}_f=u$ (Fig.~\ref{fig:vfmaxmin}) we have
$N_T=2\langle n(N_p)\rangle u t/R$ that, using (\ref{eq:stdvf}) and
recalling that $N_p=NR$, means that the measured bulk velocity will
be:
\begin{equation}
v^b_f= \frac{\langle n(N_p)\rangle}{N_p} u\,.
\label{eq:vbhighr}
\end{equation}
The above formula would give $v_f^b=u$ only if $\langle
n(N_p)\rangle=N_p$ in the bulk bins.  Therefore,
the numerical data shown in Fig.~\ref{fig:soprasoglia} provide a
strong indication that the expectation $\langle n(N_p)\rangle=N_p$ is
violated.

In Appendix~\ref{app:logistic}, for stochastic logistic kinetics
without transport, we show that the average number of individuals,
$\langle n(N_p)\rangle$, at equilibrium can be computed analytically,
see Eq.~(\ref{eq:carrying0}). In particular, when $N_p\gg 1$ we have
that $\langle n(N_p)\rangle\approx N_p-1$.  Plugging this asymptotic
expression in (\ref{eq:vbhighr}) yields the heuristic formula
(\ref{eq:vergnifit}) with $C=1$, not far from the value $C\approx
1.31$ obtained from a best fit of the numerical data.  Clearly, under
the action of transport mechanisms the number of particles in the bin
will depend not only on the reaction dynamics inside the bin but also
on the migration from and toward neighboring bins. Most likely, the
fluctuations induced by the transport process are responsible for the
deviation of $C$ from $1$.

We conclude this section noticing that similar corrections to the
front speed due to the fact that in the bulk $\langle
n(N_p)\rangle\neq N_p$ should be present also for $r\tau<1$. However,
they are much smaller than the effects discussed in the previous
section. Indeed small differences of the bulk velocity from the front
speed based on the extremal particles, when $r\tau<1$, can be detected
only for small values of $N_p$ (not shown), where they are stronger.

\section{Extension to Levy walks\label{sec:extension}}
The model presented in Sec.~\ref{sec:model} can be easily generalized
in order to account for more general transport processes, such as
L\'evy walks \cite{zaburdaev2015levy} that can model the transport
properties of several biological
populations~\cite{viswanathan1999optimizing,bartumeus2007levy,mierke2013integrin,ariel2015swarming},
simply modifying the distribution of the walk durations. For instance,
with the choice
\begin{equation}
  P(T)= (\alpha-1) T^{-\alpha} \Theta(T-1) \,,
  \label{eq:probT}
\end{equation}
at varying the value of $\alpha$ different transport processes can be
obtained.  Indeed the second moment of the displacement behaves as
\cite{andersen2000}:
\begin{equation}
\langle x^2(t)\rangle \sim 
\left\{
\begin{array}{ll}
t^{2} & 1<\alpha<2\\
t^{4-\alpha} & 2<\alpha<3\\
t & \alpha>3\,,
\end{array}
\right.
\end{equation}
i.e. it is ballistic, superdiffusive or diffusive depending on $\alpha$.
Notice that the persistent random walk previously investigated is
retrieved in the limit $\alpha\to \infty$. 
When $\alpha>3$, the diffusive motion stems from the fact that 
$\langle T^2\rangle$ is finite and according to Eq.~(\ref{eq:D})
the diffusion coefficient is equal to
\begin{equation}
  D=\frac{u^2}{2} \frac{\langle T^2\rangle}{\langle T \rangle}=\frac{u^2}{2}\frac{\alpha-2}{\alpha-3}\,.
  \label{eq:Dofalpha}
\end{equation}
However, even if the diffusion coefficient is well
defined for $\alpha>3$, this does not mean that the underlying process
is diffusive in a standard way, i.e. it is not true that $\langle
x^{2q} \rangle \sim t^{q}$ as expected for a standard diffusive process, see
\cite{andersen2000,vulpio} for a discussion. As a consequence, in the
case $\alpha<\infty$ the continuum limit of discrete stochastic 
reactive models, like ours, is nontrivial and can be
defined only in the form of an integro-differential equation with a
kernel describing the transport process
\cite{fedotov2001front,fedotov2016single,stage2016proliferating}.
However it is still possible to provide an approximate expression for
the front speed by appropriately generalizing the results of the previous
sections, when the mean square particle displacement has a diffusive
behavior (i.e. for $\alpha>3$).

Before discussing this point, let us mention that when $\alpha<3$ it
is physically reasonable to expect that $v_f \approx u$, besides
possible finite $N_p$ corrections (Sec.~\ref{sec:highrates}).
This result finds analytical support in
Ref.~\cite{fedotov2016single} in the strong ballistic case
($\alpha<2$). When the transport process is superdiffusive
($2<\alpha<3$), it should similarly hold, due to the large statistical
weight of events characterized by particles keeping their velocity for
a very long time. In both cases, tests in numerical simulations of the
discrete model confirm the expectation $v_f \approx u$ but for
finite-$N_p$ corrections of the type discussed in
Sec.~\ref{sec:highrates} (results not shown).

Let us now focus on the range $3<\alpha<\infty$, where the motion is
diffusive with diffusion coefficient given by Eq.~(\ref{eq:Dofalpha}).
In this case, the phenomenology of front propagation should not be too
different from the one described by the RCE (see
Sec. \ref{sec:cattaneo}A). In other terms we can conjecture that, when
$r$ is sufficiently small, the continuum front speed is given by $v_0
=2\sqrt{D(\alpha)r}/(1+r\tau(\alpha))$, with $D(\alpha)$ as from
Eq.~(\ref{eq:Dofalpha}) and $\tau(\alpha)=\langle
T\rangle/2=(\alpha-1)/[2(\alpha-2)]$, while $v_f=u$ for large enough
$r$.  Hence, substituting the expression of $D(\alpha)$ and
rearranging the terms, the front speed should be given by
\begin{equation} v_0=\left\{
  \begin{array}{ll}
\frac{2u\sqrt{r\tau}}{1+r\tau} 
\sqrt{\frac{\langle T^2\rangle}{\langle T\rangle^2}} & \mathrm{if} \quad r\tau<1\\
u & \mathrm{if} \quad r\tau\geq 1,
  \end{array}
  \right.
  \label{eq:velalfafinito}
\end{equation}
in close analogy with Eq. (\ref{eq:vfront}), apart from a finite $\alpha$ correction 
controlled by the ratio $\langle T^2\rangle/\langle T\rangle^2=(\alpha-2)^2/[(\alpha-3)(\alpha-1)]$ 
and the fact that now $\tau$ depends on $\alpha$.
\begin{figure}[h!]
\centering
\includegraphics[width=0.9\columnwidth]{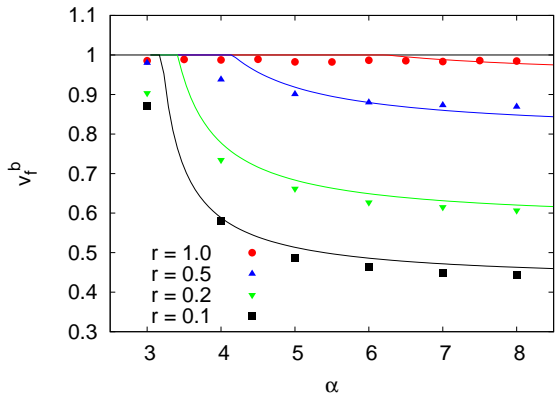}
\vspace{-.4truecm}
\caption{(Color online) Front speed vs $\alpha$ 
for various values of $r$, with $u=1$ and $N_p = 100$. Solid lines represent the approximation 
(\ref{eq:velalfafinito}). \label{fig:vsalpha}}
\end{figure}

To test the validity of prediction (\ref{eq:velalfafinito}), we
measured the bulk front speed in numerical simulations of the
stochastic particle model with the walk-duration probability density
function (\ref{eq:probT}) for several values of $\alpha$ and $r$, with
$N_p=100$ and $u=1$. The results are reported in
Fig. \ref{fig:vsalpha}, where the continuous lines represent the
prediction (\ref{eq:velalfafinito}). We can first remark that, for any
fixed $\alpha$ the front speed tends to the ballistic velocity $u$
with growing $r$ and the convergence is the faster the smaller
$\alpha$.  As for the dependency on $\alpha$ at fixed $r$, the
theoretical prediction describes fairly well the numerical data when
$\alpha$ and $r$ are such that $r\tau(\alpha)\ll 1$. The agreement
improves as $\alpha$ gets larger, which is reasonable considering that
the argument developed above amounts to a correction to the front
speed in the RCE (\ref{eq:vfront}) due to finite $\alpha$. The more
important deviations observed when $\alpha$ approaches 3 are likely
due to the increased statistical significance of persistent walks of
particularly large duration. Moreover, the case $\alpha=3$ is
marginally diffusive as $\langle x^2(t)\rangle \sim t\log(t)$.  For
$\alpha=4$ and $8$ we also studied the dependence of the front speed
on $N_p$. We found that the fluctuations induced by demographic
stochasticity have effects that are quantitatively similar to those
discussed in the case of the persistent random walk for low
(Sec.~\ref{sec:lowrates}) and high (Sec.~\ref{sec:highrates}) reaction
rates (results not shown). 
However, it is worth mentioning that for smaller values of
$\alpha$ the probability of walks lasting for a
long time increases and the assessment of the effects of
discreteness becomes more difficult, as longer simulations 
as well as averages over a larger number
of realizations are needed to safely estimate the front speed.

\section{Conclusions\label{sec:concl}}
We investigated the dynamics of a system of logistically reacting
individuals that move according to a one-dimensional persistent random
walk, focusing on front propagation and the effect of
finite-population fluctuations on it. 
Such a description of the transport process allows to remedy 
the unphysical features
(such as infinite velocities) of the standard diffusive approximation, which 
cause an overestimate of the speed of traveling waves.

After deriving the continuum limit of the individual-based model,
which corresponds to the RCE, in order to study the effects of
discreteness, we introduced a low-population-density cutoff in the
reaction term of the continuous-model equation. This allowed us to
quantify the correction to the front speed due to the finite number of
particles. For low reaction rates ($r\tau<1$) it has been possible to
analytically compute it by generalizing the treatment previously
introduced for the FKPP model \cite{brunet1997}. Similarly to that
case, we found that the correction is logarithmic with the density of
carrying capacity $N$, $v_0-v_f \sim (\log(1/N))^{-2}$ (with $v_0$ the
value from the continuum theory), in good agreement with the results
of numerical simulations of the discrete model. For high reaction
rates ($r\tau>1$), instead, the numerics indicate that the RCE is
insensitive to the cutoff. However, demographic stochasticity does
impact the particle dynamics. This result is subtle and tightly
related to the definition of the front speed. When the latter is
computed from the position of the farthest particle from the origin,
the results of the continuum are reproduced. Nevertheless, when $v_f$
is computed from the growth rate of the total number of particles, our
numerical calculations indicate that $u-v_f \sim N_p^{-1}$, where
$N_p$ is the local carrying capacity. Such a reduction of the front
speed (with respect to the ballistic velocity) hence originates from
the effect of the stochastic nature of the dynamics on the bulk
properties of the system, namely from the reduction of the effective
(average) carrying capacity, as also confirmed by a simplified
probabilistic model (developed in Appendix~\ref{app:logistic}).

While the results for the case $r\tau<1$ share important formal
similarities with the analogous ones holding for FKPP dynamics
\cite{brunet1997}, those obtained for $r\tau>1$ are more original and
specific to the RCE, and had not been documented before.  It is worth
to remark that, from a biological point of view, the latter regime
corresponds to a situation in which individuals reproduce faster than
the typical time at which they change their direction. According to
previous studies \cite{holmes1993diffusion} this condition is
difficult to achieve even by selecting organisms with high intrinsic
growth rate $r$.  Nevertheless, we believe that it might still be of
importance in the case of fast reproducing (parasites or pathogens)
species that, similarly to the spreading by long-range dispersal
considered in \cite{HF2014}, are transported by other organisms,
characterized by a highly correlated motion.

Finally, we provided an extension of the above picture for power-law
distributed walk durations, as is the case when transport is governed
by a L\'evy walk process relevant to several biological
populations~\cite{viswanathan1999optimizing,bartumeus2007levy,mierke2013integrin,ariel2015swarming}.  In particular, in the diffusive
regime ($\alpha>3$), we have shown that the front speed of reaction
fronts is well predicted by the RCE with the appropriate diffusion
coefficient, at least for not too small values of $\alpha$, and we
determined the $\alpha-$dependent correction to the asymptotic front
speed in the low-reaction-rate limit.

The predictions obtained in this work concern measurable quantities,
such as the front speed and the carrying capacity. Therefore they can
be usefully compared to experimental data. We hope that they can
stimulate experimental researches and contribute to the understanding
of the complex dynamics of biological and chemical reactive species in
realistic situations, where correlated movements represent an
unavoidable feature.

\begin{acknowledgments}
We thank R. Natalini for illuminating discussions on the numerical
integration of the RCE and S. Pigolotti for reading the manuscript 
and usefull suggestions.
\end{acknowledgments}

\appendix
\section{Calculation of the front speed for the RCE with a small cutoff\label{app:derrida}}
We consider here the RCE (\ref{eq:cattaneo2}) with a cutoff $\epsilon$
in the reaction term, i.e.  $f(n)$ is replaced by
$f(n)\Theta(n-\epsilon)$, $\Theta$ being the Heaviside step
function. Following Ref. \cite{brunet1997} we compute the corrections
to the front speed due to the cutoff, obtaining Eq.~(\ref{eq:prediction})
for the RCE.

Assuming 
$n(x,t)=h(x-v_\epsilon t)=h(z)$, Eq.~(\ref{eq:cattaneo2}) takes the form
\begin{equation}
(1\!-\!av_\epsilon^2) h^{\prime\prime} \!+\! v_\epsilon [1\!-\!af^{\prime}(h)\Theta(n\!-\!\epsilon)]\!+\!f(h)\Theta(n\!-\!\epsilon)\!=\!0\,,
\label{eq:rcecutoff}
\end{equation}
with $f(h)=h(1-h)$.
For $\epsilon\ll 1$ we can identify three regions: (I) $\epsilon \ll h\ll 1$, where the cutoff has no influence on the front; (II) $\epsilon \lesssim h\ll 1$, where the cutoff effects are important; $h<\epsilon$ where the reaction is absent. 

In region (I) the front, being unaffected by the cutoff, 
for large $z$ and small $h$, will be of the form:
\begin{equation}
h_{I}(z)\approx Aze^{-\lambda_0 z}\,,
\label{eq:hI}
\end{equation}
with $\lambda_0$ as in (\ref{eq:v0}). Indeed for $\lambda=\lambda_0$ the dispersion relation (\ref{eq:disprel}) 
attains its minimum where 
$\lambda_0$ is a degenerate root of the characteristic equation.  
In regions (II) and (III), Eq.~(\ref{eq:rcecutoff}) can be linearized as
\begin{eqnarray}
(1-av_\epsilon^2) h_{II\phantom{I}}^{\prime\prime} &+& v_\epsilon(1-a)h^{\prime}_{II}+h_{II} =0\label{eq:regionII}\,,\\
(1-av_\epsilon^2) h_{III}^{\prime\prime}&+&v_\epsilon h^{\prime}_{III}=0\,.
\label{eq:regionIII}
\end{eqnarray}
Equation~(\ref{eq:regionII}) is the same as Eq.~(\ref{eq:linear}), and
can be solved similarly by assuming $h_{II}(z)\propto
e^{-\lambda_\epsilon z}$. However, here we have an effect of the
cutoff $\epsilon$, i.e. the $\lambda_\epsilon$ solving the
characteristic equation depends on $\epsilon$. Denoting with
$0<\Delta\ll 1$ the difference $v_0-v_\epsilon$, since $v_0$
corresponds to the minimum of the dispersion relation
(\ref{eq:disprel}) we have that
\begin{equation}
v_\epsilon-v_0=-\Delta \approx (1/2) v^{\prime\prime}(\lambda_0) (\lambda_\epsilon-\lambda_0)^2\,,
\label{eq:eqlambdae}
\end{equation}
implying that we have two complex conjugate roots,
i.e. $\lambda_\epsilon=\lambda_\epsilon^r \pm i \lambda_\epsilon^i$,
and from (\ref{eq:eqlambdae}) clearly we have $\lambda_\epsilon^{i}
\sim \Delta^{1/2}$ while $\lambda_\epsilon^r\approx \lambda_0$.
Since we now have two complex conjugate roots, 
Eq.~(\ref{eq:regionII}) is solved by
\begin{equation}
h_{II}(z)\approx  C e^{-\lambda^r_\epsilon z} \sin(\lambda_\epsilon^{i} z +D)\,.
\label{eq:hII}
\end{equation}
Equation~(\ref{eq:regionIII}) instead has the solution
\begin{equation}
h_{III}(z)=\epsilon \exp[-v_\epsilon(z-z_0)/(1-av_\epsilon)]\,,
\label{eq:hIII}
\end{equation}
the front reaching the cutoff value at $z_0$.

Thus we end up with four unknowns: $C,D,z_0$ and $v_\epsilon$
(assuming $A$ as given from the unperturbed dynamics), which have to
be fixed imposing the continuity of $h$ and of its derivative at the
borders between regions I/II and II/III. It is easy to see that to
match the functions (\ref{eq:hI}) and (\ref{eq:hII}), one must require
$D=0$ so that, thanks to the fact that $\lambda_\epsilon^{i} \sim
\Delta^{1/2} \ll 1$, by expanding the sine and to leading order in
$\Delta^{1/2}$ we have $C=A/\lambda_\epsilon^i$.  Then, by imposing
the continuity of (\ref{eq:hII}) and (\ref{eq:hIII}) and of their
derivatives at $z_0$ we obtain the two relations:
\begin{eqnarray}
 A e^{-\lambda_\epsilon^r z_0} \sin(\lambda_\epsilon^i z_0)&=&\epsilon \lambda_\epsilon^i \label{eq:c1}\\
  A e^{-\lambda_\epsilon^r z_0} (-\lambda_\epsilon^r \sin(\lambda_\epsilon^i z_0)+\lambda_\epsilon^i\cos(\lambda_\epsilon^i z_0))&=& -\epsilon \lambda_\epsilon^i \frac{v_\epsilon}{(1-av_\epsilon^2)}   \,.\nonumber   
\end{eqnarray}
Dividing the second by the first yields 
\begin{equation}
-\lambda_\epsilon^r+
\frac{\lambda_\epsilon^i}{\tan(\lambda_\epsilon^i z_0)}=-\frac{v_\epsilon}{1-av_\epsilon^2}\,,
\label{eq:derrida}
\end{equation}
which is similar but not identical to that obtained by
\cite{brunet1997}. In order to fix the value of $z_0$ using the above
expression we recall that $\lambda_\epsilon^i\sim \Delta^{1/2}$ and
$\lambda_\epsilon^r\approx \lambda_0$, and to the same order
$v_\epsilon\approx v_0$. Substituting these approximations in
(\ref{eq:derrida}) and using Eq.~(\ref{eq:v0}), after simple algebra
one obtains ${\lambda_\epsilon^i}/{\lambda_0^2}\approx
-\tan(\lambda_\epsilon^i z_0)$.  The last equation can be solved by
assuming $\lambda_\epsilon^i z_0\approx \pi-\beta$ with $\beta\ll 1$
and Taylor expanding the tangent which gives
$\beta=\lambda_\epsilon^i/\lambda_0^2\propto \Delta^{1/2}$,
consistently with the assumption of a small quantity. Substituting 
$\lambda_\epsilon^i z_0\approx \pi-\beta$ in the argument of the sine
in the first of (\ref{eq:c1}), Taylor expanding and solving for $z_0$ we obtain to leading order 
for $\epsilon \ll 1$: $z_0\approx -\log(\epsilon\lambda_0^2)/\lambda_0\approx -\log\epsilon /\lambda_0$. 
Then, to order $\Delta^{1/2}$, we have 
$\lambda_\epsilon^i\approx \pi/z_0\approx -\pi\lambda_0/\log\epsilon$.  
Finally, using the above results, the fact that $\lambda_\epsilon-\lambda_0\approx i\lambda_\epsilon^i$ 
and Eq.~(\ref{eq:eqlambdae}) we obtain the result (\ref{eq:prediction}) of Sec.~\ref{sec:cutoff}, 
which was the goal of this Appendix.

\begin{figure}[b!]
\centering
\includegraphics[width=0.85\columnwidth]{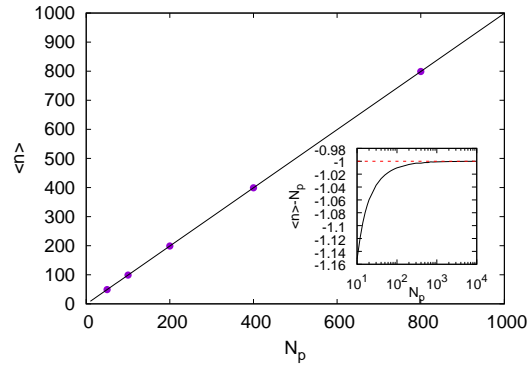}
\vspace{-.4truecm}
\caption{Average number of individuals $\langle n\rangle$ vs $N_p$ as computed by 
simulations, based on a standard Gillespie
  algorithm \protect\cite{gillespie}, of the non-spatial stochastic
  logistic model (symbols) and corresponding exact solution
  (\ref{eq:carrying0}) (solid line), obtained from
  \textit{Mathematica}. The inset shows the analytical curve $\langle
  n\rangle-N_p$ and how it converges to $-1$, which justifies
  Eq.~(\ref{eq:nasym}). \label{fig:conto}}
\end{figure}
\section{Exact solution of the stochastic logistic equation\label{app:logistic}}
In this section we consider the discrete logistic dynamics  in the absence of transport. The 
number of individuals at time $t$, $n(t)$, evolves according to the
kinetics (\ref{eq:birth}-\ref{eq:rate3}), i.e. it increases (decreases) by one
with a rate $W^+(n)=rn$ 
respectively $W^{-}(n)=rn^2/N_p$, 
$N_p$ denoting the carrying capacity).  Thus the
probability to have $n$ individuals at time $t$ evolves according to
the master equation,
\begin{eqnarray}
\partial_t P_t(n)&=& W^+(n-1) P_t(n-1)+W^-(n+1) P_t(n+1)\nonumber \\
&-&(W^{+}(n)+W^{-}(n))P_t(n)\,.
\label{eq:master}
\end{eqnarray}
At equilibrium, the detailed balance 
condition,  $P(n\!+\!1)W^-(n\!+\!1)\!=\!P(n)W^+(n)$ (where $P(n)\!=\!\lim_{t\to \infty} P_t(n)$), should hold, so that we can write
the recurrence relation
$P(n+1)=\frac{nN_p}{(n+1)^2}P(n)$,
which is solved by
\begin{equation}
P(n)=\frac{(N_p)^n}{n\cdot n!}P(1)\,,
\label{eq:pn}
\end{equation}
where $P(1)$ can be fixed using the normalization condition,
$\sum_{n=1}^{\infty}P(n)=1$. Using \textit{Mathematica} we obtained
\begin{equation}
\frac{1}{P(1)}=\sum_{n=1}^{\infty}\frac{(N_p)^n}{n\cdot n!}=
-\gamma-\Gamma(0,-N_p)-\log(-N_p)\,,
\label{eq:p1}
\end{equation}
where $\gamma$ is 
Euler-Mascheroni constant, and $\Gamma(0,-N_p)$ is
the upper incomplete Gamma function.  Once we have the expression for
$P(n)$ we can compute the average number of 
individuals, $\langle n\rangle$, at stationarity as
\begin{equation}
\langle n \rangle = \sum_{n=1}^{\infty}n P(n)=(e^{N_p} - 1)P(1)
\label{eq:carrying0}
\end{equation}
which asymptotically reaches $N_p$, but the correction for large $N_p$
goes as follows (see also Fig.~\ref{fig:conto})
\begin{equation}
\langle n \rangle\approx {N_p-1}\,.
\label{eq:nasym}
\end{equation}


%

\end{document}